\documentclass[12pt]{article}
\usepackage[margin=1in]{geometry}
\usepackage{helvet}
\usepackage{graphicx} 
\usepackage{hyperref}
\usepackage{cite} 
\usepackage{xcolor}
\usepackage{amsmath}
\usepackage{setspace}
\title{Three-Dimensional to Layered Halide Perovskites: A Parameter-Free Hybrid Functional Method for Predicting Electronic Band Gaps}

\author{$^{\dagger}$Ibrahim Buba Garba, $^{\dagger}$Lorenzo Trombini, $^{\dagger}$ Claudine Katan, \\ $^{\ddagger}$Jacky Even, $^{\ddagger}$Marios Zacharias, $^{\dagger}$Mikael Kepenekian, $^{\dagger}$George Volonakis\\
\\
\small $^{\dagger}$Univ Rennes, ENSCR, CNRS, ISCR (Institut des Sciences Chimiques de Rennes) \\
\small UMR 6226, France \\
\small $^{\ddagger}$Univ Rennes, INSA Rennes, CNRS, Institut FOTON - UMR 6082, Rennes, France}
\date{}

\begin{document}

\maketitle

\newpage
\begin{abstract}
Accurate prediction of electronic band gaps in halide perovskites using density functional theory is crucial for optoelectronic applications. Standard hybrid functionals like HSE and PBE0 are becoming computationally accessible, yet can fail at predicting the band gaps for three-dimensional (3D) and/or layered halide perovskite. This study evaluates the doubly screened dielectric-dependent hybrid (DSH) functional for predicting band gaps of Pb- and Sn-based inorganic and hybrid 3D halide perovskites, as well as layered hybrid perovskites. The DSH functional employs material-dependent mixing parameters derived from macroscopic dielectric constants, and accurately predicts band gaps for 3D perovskites only if structural local disorder is taken into account. For layered hybrid perovskites, the DSH functional based on average dielectric constants tends to overestimate the band gaps. To improve predictions, we propose using the calculated dielectric constant of the respective 3D perovskites to define the DSH screening. This method is then applied to Pb- and Sn-based layered halide perovskites with various organic spacers and multilayered structures, such as BA$_2$MA$_{n-1}$Pb$_n$I$_{3n+1}$ with n =1, 2, 3, resulting in improved precision. The HSE functional systematically underestimates band gaps in layered perovskites due to the missing non-local long-range dielectric screening. On the other hand, the PBE0 is in good agreement with the experimental values, in particular for the layered iodide perovskites. The computational framework introduced here provides an efficient parameter-free \textit{ab initio}  methodology suitable for predicting the electronic properties of 3D, layered halide perovskites and their heterostructures, towards modelling materials for advanced optoelectronic devices.
\end{abstract}

\newpage

Lead-based hybrid halide perovskites have emerged as versatile materials with tunable structural, electronic, and optical properties due to their unique inorganic corner-sharing octahedral lattices and complex organic components. These diverse material compositions enable a spectrum of functionalities that have been successfully applied in various technological fields, such as solar cells, photodetectors, and light-emitting devices \cite{Peng_2023, Li_2021, Katan_2019, Han_2022}. Halide perovskites are generally classified into three-dimensional (3D) and layered halide perovskites. The 3D halide perovskites are represented by the general formula AMX$_3$, where M is typically a divalent cation in the center of the octahedron such as Pb$^{2+}$ or Sn$^{2+}$, X is a halogen ( I, Br, or Cl), and the A site is typically occupied by Cs$^{+}$ or organic cations such as methylammonium (MA) or formamidinium (FA). In recent years, solar cells based on 3D halide perovskites have achieved remarkable power conversion efficiencies today comparable to those achieved by silicon cells~\cite{Peng_2023}. However, these 3D materials can suffer from poor environmental stability when exposed to air and/or illumination~\cite{An_2021}. Their layered counterparts, also called two-dimensional (2D) halide perovskites, are further classified into Ruddlesden-Popper (RP), Dion-Jacobson (DJ), and alternating cation (ACI) structures, depending on the choice of organic molecules that separate the layers and the resulting stacking arrangement~\cite{Blancon_2020}. For example, the RP structure typically has two monovalent organic cations and the general formula $A_2A'_{n-1}M_{n}X_{3n-1}$, where $n$ is the number of connected MX$_6$ octahedra in the structure~\cite{Mitzi_1995}. Layered halide perovskites can exhibit significantly higher material stability than 3D perovskites. However, their photovoltaic performance is usually lower~\cite{Li_2022, Cucco_2024}. Within these layered materials, the organic moieties are usually insulating; thus, the charge carriers are primarily constrained to the inorganic layers. This leads to the natural formation of quantum wells (QW), in which the optical band gap increases with decreasing $n$ $(n=1,2,3,...)$ due to quantum and dielectric confinement~\cite{Blancon_2018, Gedda_2021}. Most recently, the combination of the two in the so-called 3D/2D heterostructures has achieved superior stability compared to 3D and higher efficiency than layered halide perovskites, making them a promising platform for optimizing device performance in various architectures~\cite{Metcalf_2023}. Consequently, knowledge about the energy levels is critical to monitor the band alignment between layered and 3D structures~\cite{Metcalf_2023}.

Density Functional Theory (DFT) calculations, employing semi-local exchange-correlation (XC) functionals like the generalized gradient approximation (GGA) with spin-orbit coupling effects, lead to a sizable and systematic underestimation of the electronic band gap. The state-of-the-art method to go beyond DFT is Hedin's $GW$ many-body quasiparticle perturbation theory~\cite{Hedin_1965} that accounts for exchange-correlation effects via the self-energy, defined based on the Green's function, $G$, and the dynamical Coulomb screening potential $W$. The $GW$ quasiparticle conduction band minimum (CBM) and valence band maximum (VBM) can be interpreted as the ionization energy (IE) measured by ultraviolet photoemission spectroscopy (UPS) and the electron affinity (EA) obtained from inverse photoemission spectroscopy (IPES), respectively. However, the $GW$ approximation is computationally expensive~\cite{Umari_2014}, which limits its use in large systems containing hundreds or thousands of electrons, such as layered halide perovskites and very few works have reported the GW band gaps in the entire structure (i.e., including the organic moieties) of layered halide perovskites~\cite{Yin_2017, Giorgi_2018, Molina-Sánchez_2018, Filip_2022, Leppert_2024}.

An alternative approach to $GW$ is using the so-called hybrid functionals within DFT constructed by admixing a non-local Hartree fraction to the local exchange potential~\cite{Martin_2020}. This mixing mitigates the self-interaction error, and has been shown for several semiconductors to result in improved band gaps and reasonable energy level alignment at a computational cost that can be significantly smaller than $GW$ calculations~\cite{Chen_2014}. Well-known hybrid functionals include PBE0~\cite{Perdew_1996}, which adds 25\% (i.e., the mixing parameter $\alpha_{H}$ is set to 0.25) exact Hartree-Fock exchange to the exchange-correlation potential. HSE~\cite{Heyd_2003, Krukau_2006} is similar to PBE0 but introduces a range-separation parameter $\mu$ that partition the Coulomb interaction into short- and long-range components. These hybrid functionals significantly improve standard DFT, yet PBE0 tends to overestimate band gaps, particularly for narrow band gap materials. At the same time, HSE underestimates band gaps of moderate band gap semiconductors and wide band insulators~\cite{Steiner_2014}. To overcome this limitation, the mixing parameters $\alpha_{H}$ and $\mu$ are often modified to reproduce experimental results, which limits the predictive capability of hybrid functionals~\cite{He_2012, Du_2014, Traore_2022}. A good example of this limitation is the case of halide perovskites, for which hybrid functionals are being employed with a wide range of mixing parameters, in many cases surpassing the 60\% of exact exchange ~\cite{Du_2014, Basera_2021, Traore_2022, Li_2020}. Such manual tuning of the parameters in hybrid functionals is impractical and ineffective if the band gap of the material is not known \textit{a priori}. It can also be challenging to predict suitable mixing parameters in complex heterostructures. Hence, it is desirable to develop mixing parameters independent of empirical tuning and have an intrinsic physical meaning.

To this end, it has been shown that the inverse macroscopic dielectric constant $\epsilon^{-1}_{\infty}$, which relates to the screening strength in materials, can be employed in a "parameter-free" dielectric-dependent hybrid functional (DDH) which gives good agreement with experiments for representative materials with a wide range of band gaps and dielectric constants ~\cite{Alkauskas_2011,Marques_2011,Moussa_2012}. A fully self-consistent version sc-DDH, which involves evaluating $\epsilon_{\infty}$ at the end of each DDH cycle, was also developed~\cite{Skone_2014}. A range-separated version (sc-RS-DDH)~\cite{Skone_2016}, with the long-range screening set at the empirical value of 0.25, has been proposed that further improves the band gap prediction with respect to sc-DDH. In the same spirit, but without any empirical parameter, two types of dielectric-dependent hybrid functionals have recently been introduced. First, the doubly screened dielectric-dependent hybrid (DSH) functional~\cite{Cui_2018}, which accounts for both electronic and metallic screening, and the dielectric-dependent range-separated hybrid functional using the Coulomb attenuated method (DD-RS-CAM)~\cite{Chen_2018}. Although both functionals use the same model dielectric function~\cite{Cappellini_1993, Shimazaki_2008}, in each case the range separation parameter $\mu$ is determined slightly differently. In DSH, $\mu$ is derived from modified Thomas-Fermi screening, whereas in DD-RS-CAM $\mu$ is obtained by fitting the first-principles RPA dielectric function with the model dielectric function. Nevertheless, these two functionals are practically the same and do not give rise to significant differences in the predicted band gaps of materials~\cite{Skone_2016, Yang_2023}. The model macroscopic dielectric function in DSH and DD-RS-CAM disregards both local field effects (i.e., off-diagonal elements) and dynamical effects and can be viewed as a static approximation to the diagonal Coulomb hole and screened exchange (COHSEX)~\cite{Hedin_1965, Gygi_1989} approximation to the GW self-energy. Overall, the validity of this approximation depends on how well the semi-local correlation represents the Coulomb hole and the significance of the dynamical effects. \cite{Chen_2018} Table~S1 summarizes the parameters of the different hybrid functionals.

To date, DSH has been applied to 3D halide perovskites~\cite{Bischoff_2019, Wang_2023, Yang_2023, Wang_2024}, and, to the best of our knowledge, no reports are validating DSH for the prediction of the band gap of layered halide perovskites, which remains a challenge. Nevertheless, the computational studies on 3D perovskites employ high-symmetry structures and disregard effects arising from structural local disorder, which are critical for the description of halide perovskites, playing an important role on their structural, electronic, optical, and phonon properties~\cite{Zhang_2020, Zacharias_2023a, Zacharias_2023b}. In this work, we compute the electronic band gaps of the most prominent 3D and layered halide perovskites using the DSH hybrid functional. We do so to explore the possibility of developing a consistent workflow that would permit calculating the band gaps of all these materials without having to tune any parameter of the hybrid functionals empirically. To this end, we consider a set of layered halide perovskites, known to pose a challenge for hybrid functionals, containing different organic cation spacers and different halogen atoms (i.e., Cl, Br, and I), as well as a layered perovskite series $BA_2(MA)_{n-1}Pb_nI_{3n+1}$ with n = 2 and n=3. All materials studied in this work are shown in Figure~1. We assess the performance of the DSH functional in both its one-shot (DSH0) and self-consistent (sc-DSH) forms, comparing the results with those obtained using typical hybrid functionals, such as HSE and PBE0, and experimental measurements. We also explore the importance of the hybrid functionals' short- and long-range separation on band gap for each system. 

\begin{figure*}[!t]
\includegraphics[width=\textwidth]{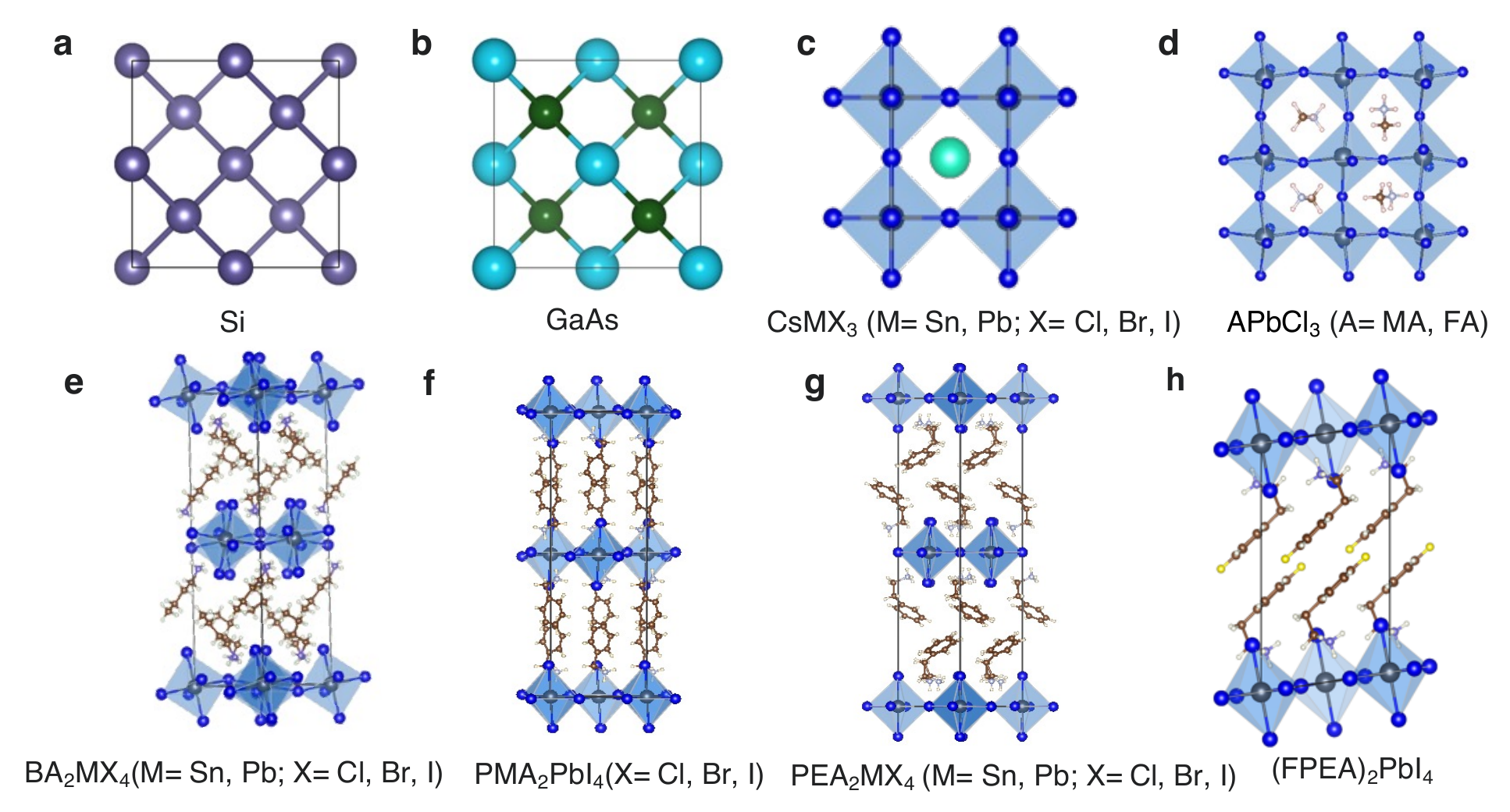}
\caption{\label{fig:3D_2D_materials} Polyhedral models of various materials with lattice parameters in Table S2 and S3: (a) Si; (b) GaAs; (c) 3D CsMX$_3$ (M = Sn, Pb; X = Cl, Br, I) inorganic perovskites; (d) 3D APbCl$_3$ (A = MA, FA) organic-inorganic perovskites (e--h)  2D layered halide perovskites with different organic cations and inorganic metal framework (M= Pb; Sn) including (e) BA$_2$MX$_4$, (f) PMA$_2$PbX$_4$,  (g) PEA$_2$MX$_4$, and (h) FPEA$_2$PbI$_4$.} 
\end{figure*}

\begin{figure*}[!t] 
\includegraphics[width=\textwidth]{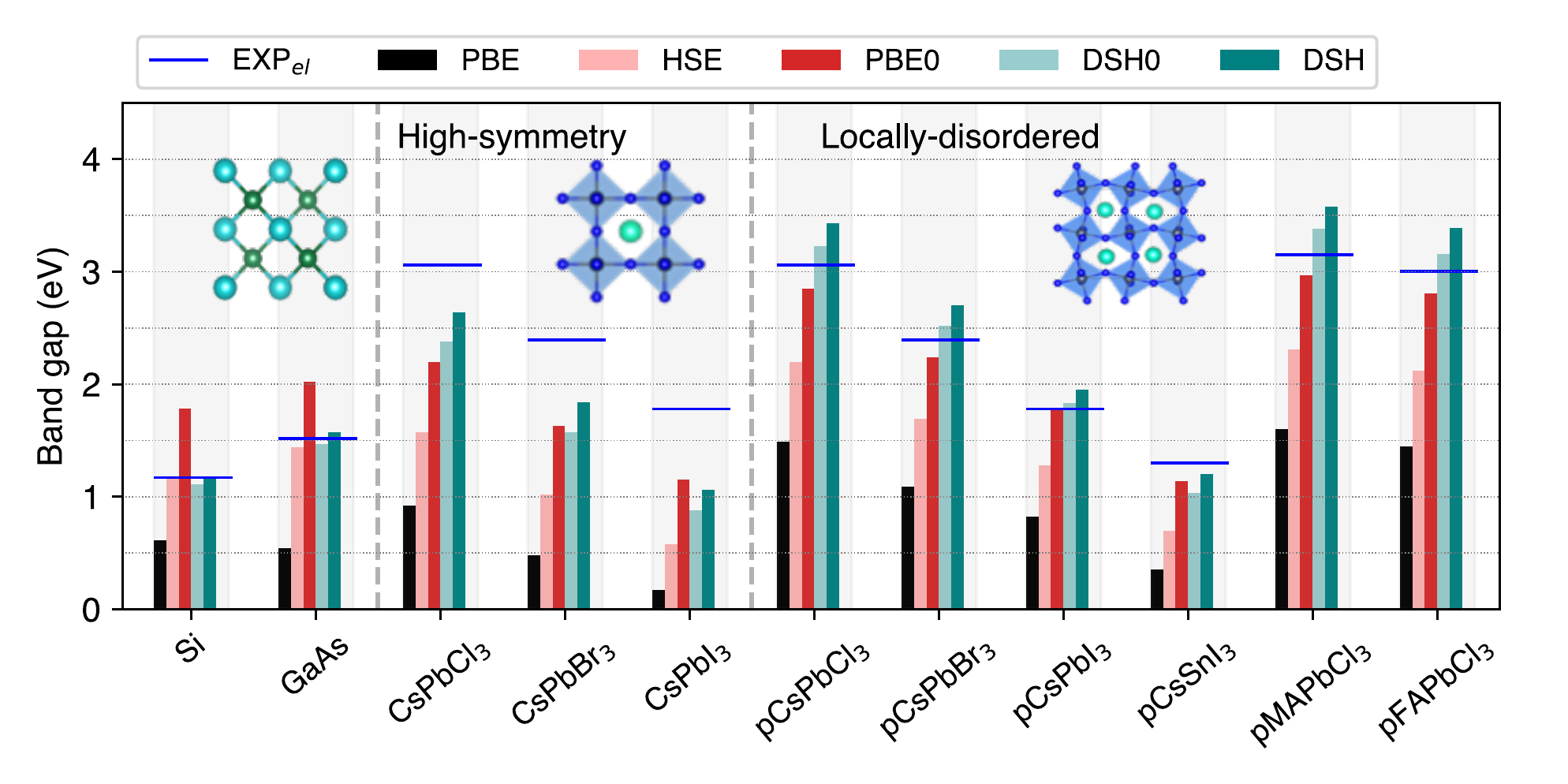}
\caption{Band gaps of three-dimensional bulk semiconductors calculated using PBE (black), HSE (salmon), PBE0 (red), DSH$0$ (light green), and DSH (dark green). The electronic experimental band gaps are represented by the blue horizontal line (See Table S2 for values). The significant enhancement of the band gap observed in p-AMX$_3$ across all the functionals is primarily due to structural local disorder within the material. }
\label{fig:3D_}
\end{figure*}

We begin by analyzing how the doubly screened hybrid (DSH) functional performs on the prediction of the band gaps of prototypical cubic structures of 3D halide perovskites AMX$_3$ (A = Cs, MA, FA; M = Sn, Pb; X = Cl, Br, I). We model the structures using the lattice parameters that are summarized in Table~S2. All computed band gaps include spin-orbit coupling effects, important in halide perovskites~\cite{Even_2013}. The structures of the organic-inorganic halide perovskites MAPbCl$_3$ and FAPbCl$_3$ are taken from Ref.~\citenum{Quarti_2025} and were generated using the ASDM method.~\cite{Zacharias_2023a, Zacharias_2023b}. We compare these predictions with those obtained using PBE, PBE0, and HSE06, as well as the measured experimental electronic band gaps (See Table S2). Figure~\ref{fig:3D_} shows the band gaps calculated employing different functionals (bars) and the measured band gaps (solid lines). As an additional reference, we include the standard semiconductors  Si and GaAs, which have similar band gaps yet dielectric constants much higher than AMX$_3$. Not surprisingly, all hybrid functionals drastically improve the PBE band gap values. For the standard semiconductors, HSE values are very close to the experimental band gaps, and so are the values obtained from the dielectric-dependent DSH0 and DSH. On the other hand, PBE0 significantly overestimates the band gap as range separation is critical for the functional form in both Si and GaAs. Moving on to the cubic phases of the 3D halide perovskites AMX$_3$, the hybrid functionals fail to describe the band gaps as all functionals significantly underestimate the measured band gaps. However, this underestimation is related to neglecting local disorder (also referred to as positional "\textit{polymorphism}"), which is critical for the accurate description of the electronic structure of halide perovskites at higher temperatures, including their dielectric properties~\cite{Zhao_2020, Zhang_2020, Zacharias_2023a, Zacharias_2023b}. Some of us have previously shown that such local disorder (shown in the inset structures of Figure~\ref{fig:3D_}) has significant effects on the band gaps, lattice dynamics, and electron-phonon coupling of halide perovskites~\cite{Zacharias_2023a, Zacharias_2023b}. Therefore, we consider the locally disordered configurations p-AMX$_3$ to assess the performance of the hybrid functionals. As shown in the last panel of Figure~\ref{fig:3D_}, we find a significant increase of the band gap values in p-CsPbX3 resulting from the local disorder for all the functionals, ranging from 570~meV (PBE, p-CsPbCl$_3$) to 960~meV (DSH0, p-CsPbI$_3$), leading to a better agreement with the measurements.

For halide perovskites, the HSE functional underestimates the band gaps and underperforms with respect to PBE0. This important difference can be explained by looking into the details of the relationship between the screening and mixing of the parameters of functionals. HSE and PBE0 employ the same non-local short-range dielectric screening, but PBE0 includes 25\% of long-range dielectric screening. This implies that for halide perovskites, the long-range dielectric screening needs to be close to the 25\% or $\epsilon_{\infty}^{-1}$, like in the case of dielectric-dependent functionals. Furthermore, we note the relatively close band gap values between PBE0 and DSH for CsPbI$_3$. As the two functionals have comparable long-range dielectric screening effects ($\alpha_{H}$ = 0.25 for PBE0 and 0.22 for DSH), yet very different short-range screening (0.25 for PBE0 and the full exact-exchange for DSH), range separation is not critical. The exact form of the screening for the materials is shown in Figure S1. When moving to lighter halogen atoms, the long-range dielectric screening fraction increases from 0.28 for p-CsPbBr$_3$ to 0.31 for p-CsPbCl$_3$, hence the DSH band gap values gradually diverge from PBE0. Overall, it is evident that for 3D halide perovskites, using the standard PBE0 or the DSH functional with polymorphous structures provides an efficient methodology for predicting band gaps without needing empirical tuning of the mixing fractions.

Next, we discuss the effects of self-consistency on improving DSH band gaps. In both full-range ~\cite{Skone_2014} and range-separated dielectric dependent hybrid functionals~\cite{Cui_2018, Chen_2018}, early results indicated that self-consistency improves the band gaps values compared to one-shot methods, especially for narrow band gap materials~\cite{Cui_2018}. However, such self-consistency worsens the results for transition metal oxides, which is attributed to the underestimated dielectric constants~\cite{Peitao_2020}. Here, we observe that self-consistency leads to relatively minor changes in the calculated band gaps, with the change being marginal in the case of Si, GaAs, and p-CsPbI$_3$. Figure S3 shows the band gap variation and dielectric constants for each DSH iteration. We also note that the method employed to evaluate $\epsilon_{\infty}$ in the self-consistency cycle can be critical. $\epsilon_{\infty}$ can be computed using different methods, including density functional perturbation theory (DFPT)~\cite{Baroni_2001}, finite electric field in the Berry phase approximation~\cite{Resta_1992}, or linear optics in the Random Phase Approximation (RPA)~\cite{GajDOS_2006}, each with different advantages and limitations discussed in the supporting information file. Here, we employ the finite electric field method, as detailed in the computational details, to include local field effects, which are important for halide perovskites. However, converging $\epsilon_{\infty}$ remains challenging when employing hybrid functionals.

\begin{figure*}[!t] 
  \includegraphics[width=\textwidth]{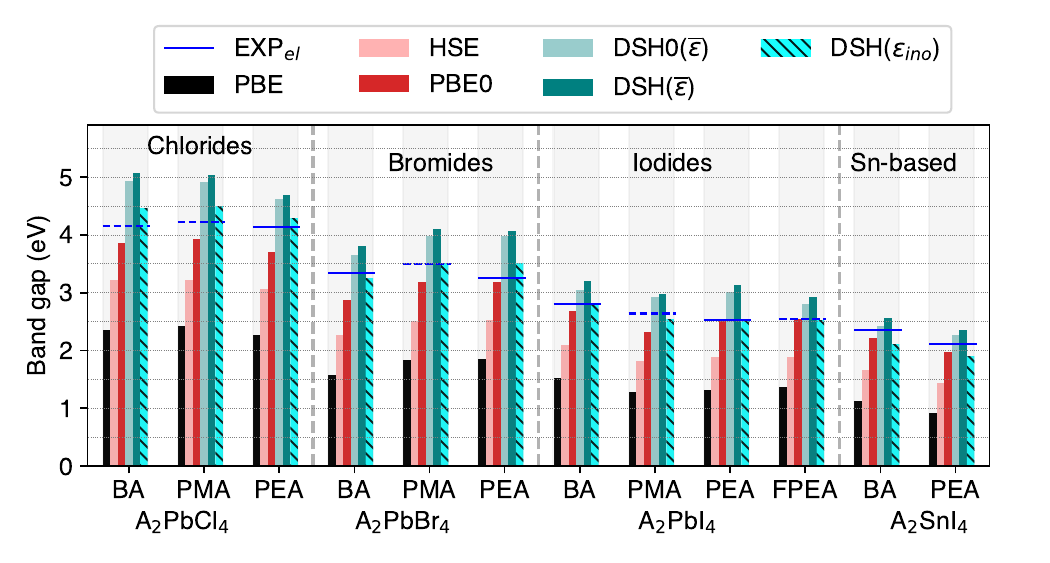}
  \caption{Comparison of band gaps in layered  halide perovskites using PBE (grey), HSE (salmon), PBE0 (red), DSH0 (light-green), DSH($\overline{\epsilon}$) (dark-green) and DSH($\epsilon_{\text{ino}}$) (hatched turquoise). 
  The electronic experimental band gap values (EXP$_{el}$) (see Table S3), are indicated by the blue horizontal lines, and dotted blue lines are derived by adjusting the optical gaps $E^{op}_g$, with exciton binding energy $E_b$, where $E_b = 0.18E^{el}_g -0.249$\cite{Hansen_2023}.}
  \label{fig:2D_band_gaps}
\end{figure*}

Next, we employ the DSH functional to set up a parameter-free workflow for layered halide perovskite materials. To this end, we consider a series of prototypical lead-based layered structures with various halogen atoms (i.e., Cl, Br, I) and different organic spacers, as seen in Table S3. Figure \ref{fig:3D_2D_materials}d-g shows the materials with the organic cations BA (butylammonium), PMA (phenylmethyl ammonium), PEA (phenylethylammonium) and FPEA (4F-phenylethylammonium), respectively.  We have not considered locally disordered structures for the layered perovskites, as our calculations are based on experimental low-symmetry structures with significant distortion and therefore, the effects are expected to be small.  For example,  the change in total energy due to structural local disorder for orthorhombic CsPbI$_3$ is reported at just 1 meV per formula unit ~\cite{Zacharias_2023a}. Also, in the absence of experimental electronic band gaps, we correct the experimental optical band gaps (E$^{\text{op}}_g$) using the empirical relation of the exciton binding energy ($E_b$), derived for layered perovskites (i.e., $E_b = 0.18E^{\text{el}}_g - 0.249$). \cite{Hansen_2023} The \textit{ab initio} computation of the exciton binding energy would require to solve the GW plus Bethe-Salpeter equation (GW+BSE). To date, only a handful of studies have used the GW+BSE method to compute band gap and exciton binding energy in the full structure (i.e., including organic moieties) of layered halide perovskites ~\cite{Yin_2017, Giorgi_2018, Molina-Sánchez_2018, Filip_2022, Leppert_2024}. For example, Yin et al.~\cite{Yin_2017} investigated the origin of light emission in two layered perovskites. Giorgi et al.~\cite{Giorgi_2018} calculated the optoelectronic properties of isolated and layered $BA_2(MA)_{n-1}Pb_nI_{3n+1}$ with n=1 and n=2. Filip et al.~\cite{Filip_2022} analyzed the effects of various organic cations on the band gaps, dielectric screening, and exciton binding energies of a set of A$_2$PbBr$_4$ compounds.

\begin{figure*}[!t]
  \includegraphics[width=0.99\textwidth]{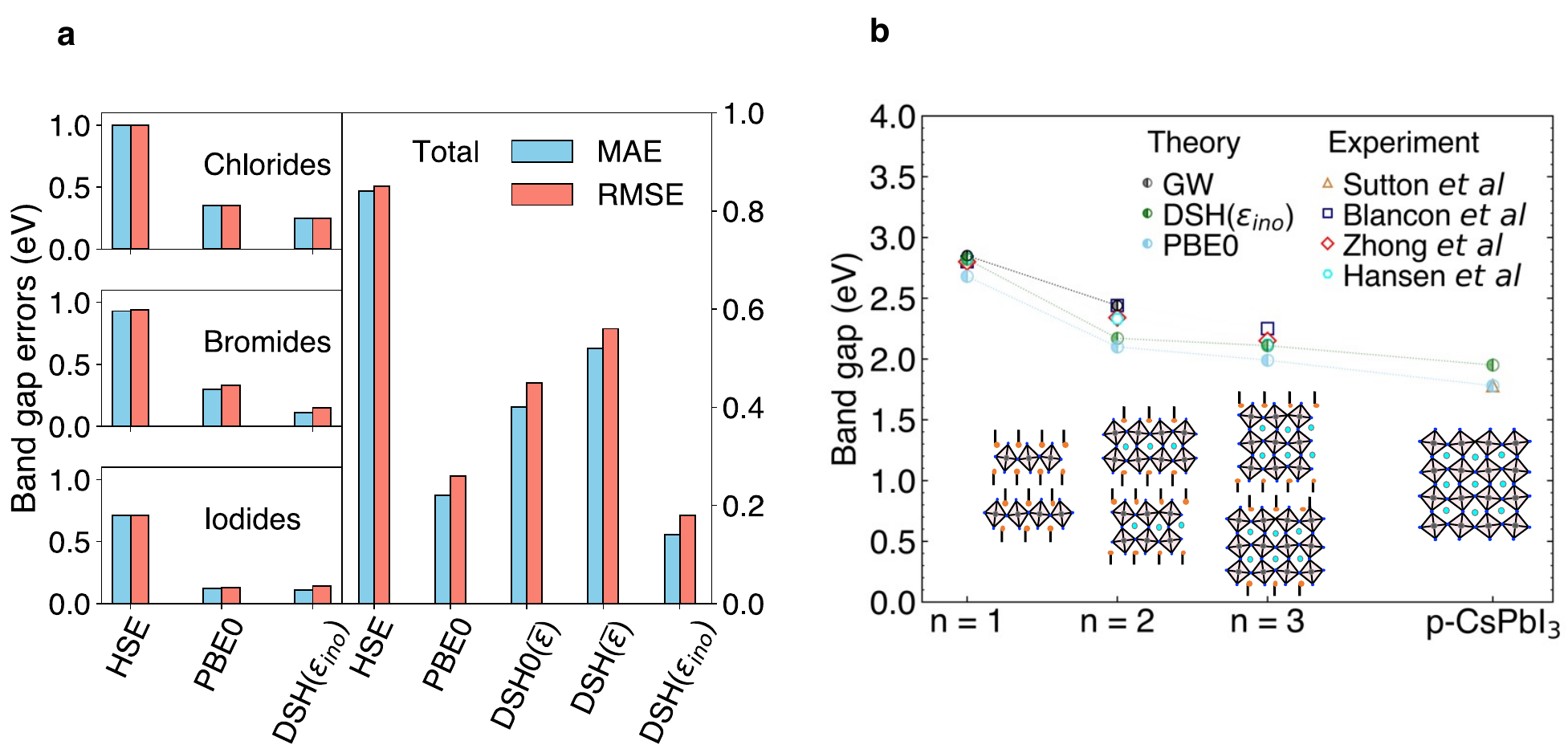}
  \caption{ (a) Comparison of Mean Absolute Error (MAE) and Root Mean Square Error (RMSE) of HSE, PBE0, DSH0($\overline{\epsilon}$), DSH($\overline{\epsilon}$), and DSH($\epsilon_{\text{ino}}$) functionals with respect to experimental electronic band gaps of 2D layered halide perovskites  grouped by the same halogen in left and total in right.  
  (b) Electronic band gaps of the layered $BA_2(MA)_{n-1}Pb_nI_{3n+1}$ series ($n = 1, 2, 3$) and  3D (p-CsPbI$_3$), computed using PBE0 (skye-blue circles) and DSH($\epsilon_{\text{ino}}$) (green circles) compared with GW calculation~\cite{Giorgi_2018} (black circles) and experimental values from Blancon \emph{et al}~\cite{Blancon_2018}, Zhong \emph{et al}~\cite{Zhong_2022}, Hansen \emph{et al}~\cite{Hansen_2023} for $n = 1, 2, 3$, and Sutton \emph{et al}~\cite{Sutton_2018} for the 3D case (See Table S4)}
  \label{fig:2D_BA_series} 
\end{figure*}

For the DSH band gap calculation, assigning a single value to the dielectric screening ($\epsilon_\infty^{-1}$) for any layered perovskite is particularly challenging as the materials are composed of layers of moieties (i.e., organic and inorganic) with different dielectric constants, hence are not homogeneous. The first, most trivial option is to use the average of the main components of the dielectric tensor (i.e., the average of the diagonal) to define the Hartree exact exchange fraction, that is $\alpha_{H}= (\overline{\epsilon})^{-1} $ with  $\overline{\epsilon}=(\epsilon_{xx}+\epsilon_{yy}+\epsilon_{zz})/3$. The calculated band gaps within the one-shot DSH$0$($\overline{\epsilon}$) and self-consistent DSH($\overline{\epsilon}$) method for layered perovskites are shown in Figure~\ref{fig:2D_band_gaps}. Similar to 3D CsPbX$_3$, the band gaps calculated employing the PBE and HSE functionals are largely underestimated compared to experiment. The band gap increases from PBE to HSE, PBE0, DSH$0$($\overline{\epsilon}$) and finally DSH($\overline{\epsilon}$). However, we find that both DSH$0$($\overline{\epsilon}$) and DSH($\overline{\epsilon}$) functionals overestimate the band gaps across all the layered materials irrespective of the crystal symmetry, organic cation, or the type of halogen in the perovskite layer. Taking for example BA$_2$PbBr$_4$ (room temperature phase), DSH0($\overline{\epsilon}$) predicts a band gap of 3.65 eV, which is 0.31 eV higher than the experimental value of 3.34 eV reported by Silver et al.~\cite{Silver_2018} The self-consistent DSH method, DSH($\overline{\epsilon}$), yields an even higher value of 3.79 eV, exceeding the experimental result by 0.45 eV. This trend is consistent across all layered materials studied here and can be explained as the screening for the DSH0 and DSH is based on $\overline{\epsilon}$, which includes contributions from the inorganic perovskite and the organic layer. In other words, the dielectric screening is an average of a low dielectric constant of the organic layer and a higher one of the inorganic layer. However, for the layered halide perovskites of interest here, the electronic band edges that define the electronic band gap are composed of states of the inorganic layer, as the valence band is mainly Pb 6$s$-orbitals and the halogen atoms $p$-orbitals, while the conduction band minimum is mostly of Pb 6$p$-orbitals and the halogens $p$ orbitals~\cite{Du_2017, Tanaka_2003}. Therefore, one can argue that using DSH with the dielectric constant solely of the inorganic perovskite layer is more reasonable for predicting the band gaps accurately. Figures S4 and S5 show the DSH($\varepsilon_{ino}$) density of states (DOS) for all layered perovskites presented in Figure~\ref{fig:2D_band_gaps} and Figure~\ref{fig:2D_BA_series}a, respectively, showing that the band edges are primarily composed of orbitals from the inorganic framework.

Considering the cubic phases of 3D halide perovskites allows defining unambiguously $\epsilon_{\infty}$. Yet, for layered halide perovskites there are various methods developed to extract the dielectric constants of the inorganic ($\epsilon_{\text{ino}}$) and organic ($\epsilon_{\text{org}}$) layers. For example, $\epsilon_{\text{ino}}$ can be approximated as the bulk dielectric constant of the respective halide salt PbI$_2$~\cite{Ishihara_1990}. One can employ a dielectric profile calculated using different flavors of DFT to assess $\epsilon_{\text{ino}}$ and $\epsilon_{\text{org}}$~\cite{Sapori_2016, Traore_2018}, or model the structure using a two capacitor model and fit the dielectric constants along one~\cite{Hansen_2023} or both directions~\cite{Sio_2022}. While these methods can provide reasonable estimates of the dielectric constants, the challenge remains in accurately defining the model's parameters, for example, the thickness of the inorganic layers or defining a clean interface between the two layers. Moreover, the accuracy of the dielectric constant depends on the choice of the DFT functional; the semi-local PBE functional, for example, tends to overestimate dielectric constants and gives divergent values for materials with narrow band gaps~\cite{Cui_2018}. To overcome this issue in a simple way that is easy to implement in an \textit{ab initio} parameter-free workflow, we propose employing the DSH using the dielectric constant of the 3D polymorphous perovskites to set the screening parameter. This method corresponds to the band gaps denoted as DSH($\epsilon_{\text{ino}}$) shown in  Figure~\ref{fig:2D_band_gaps} and summarized in Table S3. The band gap values predicted within DSH($\epsilon_{\text{ino}}$) is a significant improvement with respect to the other functionals. For example, the band gap of BA$_2$PbBr$_4$ using DSH($\epsilon_{\text{ino}}$) is 3.26 eV, in excellent agreement with the experimental value of 3.34 eV \cite{Silver_2018}. Interestingly, PBE0 also gives band gaps close to the measured electronic band gaps and similar to DSH($\epsilon_{\text{ino}}$). For the iodides in particular, the similar performance of PBE0 and DSH($\epsilon_{\text{ino}}$) can be attributed to a similar Hartree fraction for the long-range screening, which is $\alpha_H = 0.25$ in PBE0, and $\alpha_H = 0.22$ in DSH($\epsilon_{\text{ino}}$) as $\epsilon_{ino}$ is 4.52, and the effects from the short-range screening is less important as the case of 3D halide perovskites shown in Figure~\ref{fig:3D_}. Similarly, for bromides, DSH($\epsilon_{\text{ino}}$) is slightly above the measured band gaps, and PBE0 slightly below, which can also be attributed to the different screening at the long-range as $\epsilon_{\text{ino}} = 3.59$, yields $\alpha_H$ of 0.28. In contrast, for chlorides an $\epsilon_{\text{ino}} = 3.2$ gives an $\alpha_H$ of 0.31. The exact shape of the screening for all layered perovskites studied here is shown in Figure~S2. 

In order to quantify the performance of the different functionals in describing the band gap of layered halide perovskites, we calculate the mean absolute error (MAE) and root mean square error (RMSE). Figure~\ref{fig:2D_BA_series}a summarizes the performance of all hybrid functionals for all the layered materials, as well as by halogen (A$_2$PbCl$_4$, A$_2$PbBr$_4$, A$_2$MI$_4$ (M=Sn,Pb)). The HSE functional is the least suitable, with errors close to 1.0~eV. PBE0, on the other hand, predicts the band gap reasonably well across the layered materials, particularly for iodides. 

The proposed DSH($\varepsilon_{\text{ino}}$) functional achieves the highest accuracy, yielding the lowest MAE (RMSE) of 0.14 eV (0.18 eV). For all iodides, PBE0 exhibits the same accuracy, which can be practical for future calculations as one does not need to first calculate $\varepsilon_{ino}$. However, for perovskites with higher band gaps (i.e., bromides and chlorides) DSH outperforms in terms of accuracy. Overall, PBE0, DSH0($\overline{\varepsilon}$), DSH($\overline{\varepsilon}$), and  HSE, exhibit MAE (RMSE) values of 0.22 eV (0.26 eV), 0.40 eV (0.45 eV), 0.52 eV (0.56 eV), and 0.84 eV (0.85 eV), respectively.
We also include two Dion-Jacobson layered halide perovskites (see Table S3), further validating the observed performance trend for all layered materials. Since in DSH($\epsilon_{\text{ino}}$), a fixed Hartree fraction is adopted to set the screening in layered perovskites with the same halogen. Consequently, any variation of the band gap of materials with the same inorganic layer can be attributed to structural distortions of the PbX$_6$ octahedra due to the different organic cations. For instance, PEA$_2$PbBr$_4$ and PMA$_2$PbBr$_4$ have similar Pb-X-Pb bond angles of 151$^{\circ}$ and 150$^{\circ}$ respectively, resulting in practically identical band gaps of 3.50 eV. While BA$_2$PbBr$_4$, which exhibits slightly less octahedral distortion with the  Pb-X-Pb angle amounting to 155$^{\circ}$, the band gap is narrower 3.26 eV as the orbital overlap increases. A similar trend also applies to the iodides. This relationship between the increase in Pb-X-Pb distortion with the increase in the band gap is well documented and understood both theoretically~\cite{Katan_2019} and experimentally \cite{Knutson_2005}.

Finally, since the DSH($\epsilon_{\text{ino}}$) is successfully predicting the band gaps of both 3D and the different types of layered halide perovskites, we extend our analysis to verify that this approach can also predict band gaps for layered materials with a thicker inorganic layer (i.e., larger $n$-numbers). Thus, we apply our method to $BA_2(MA)_{n-1}Pb_nI_{3n+1}$, where $n$ is the number of octahedrons in the inorganic layer. We find that the band gap decreases as $n$ increases, which is expected due to the reduction of the quantum, and dielectric confinement~\cite{Stoumpos_2016, Blancon_2018, Katan_2019}. We also note that to avoid introducing an artificial dipole due to the specific orientation of the organic molecule in the perovskite lattice, for $n = 2$ and $n = 3$, we replace $MA$ in $BA_2(MA)_{n-1}Pb_nI_{3n+1}$ with Cs (see the inset of Figure~\ref{fig:2D_BA_series}). The calculated band gaps employing PBE0 and DSH($\epsilon_{\text{ino}}$) are compared to experimental electronic band gaps shown in Figure~\ref{fig:2D_BA_series}. For the 3D case, we employ the DSH experimental band gaps of p-CsPbI$_3$. As shown in Figure \ref{fig:2D_BA_series}, the DSH($\epsilon_{\text{ino}}$) band gaps agree well with the experiments for the complete series of $n= 1, 2, 3$. Figure S5 shows the DOS for all BA compounds. Similar to the other layered materials, the band edges are made of orbital contributions from the inorganic framework.
Notably, our band gap for $ n=1$ is comparable to the GW band gap reported by Giorgi et al.~\cite{Giorgi_2018} and the experimental values reported in Table S4. This comparison also highlights the predictive accuracy of DSH($\epsilon_{ino})$. It offers a useful framework that can be used for other layered halide perovskites and their heterostructures with other 3D materials. 

Overall, we present results for the band gaps of 3D and layered halide perovskite materials employing a consistent workflow to optimally set up the doubly screened dielectric-dependent hybrid (DSH) functional. 
Our results demonstrate that the proposed DSH functional, which incorporates only material-dependent mixing parameters based on macroscopic high-frequency dielectric constants, predicts band gaps more accurately than standard hybrid functionals for both 3D and layered perovskites. Our findings highlight the importance of accounting for local disorder when benchmarking any functional for 3D halide perovskites. For layered materials, it is evident that special attention must be given when setting the screening based on the dielectric constant since using the average of the diagonal components of the dielectric constant matrix in DSH0($\overline{\varepsilon}$) and DSH($\overline{\varepsilon}$), tends to overestimate the measured band gaps. Consequently, employing the DSH functional with dielectric constants from locally disordered 3D perovskite structures DSH($\varepsilon_{ino}$) to predict band gaps gives the highest accuracy with MAE of 0.14 eV across all layered perovskites, regardless of the organic spacer type, or the halogen in the inorganic framework. In contrast, the HSE functional significantly underestimates band gaps due to the lack of non-local long-range dielectric screening, resulting in the highest MAE of 0.84 eV for all materials.
On the other hand, the PBE0 functional that shares the same non-local short-range screening with HSE but has an additional 25\% non-local long-range screening performs very well, with MAE of 0.22~eV. It could be the functional of choice if one cannot afford to compute the dielectric properties of halide perovskites. The computational strategy presented here sets up a parameter-free \textit{ab initio} methodology that allows efficient prediction of the electronic properties of both 3D and layered halide perovskites. As such, the proposed workflow and benchmark can be extremely powerful in investigating 3D/2D heterostructures, which are challenging to describe at the \emph{ab initio} level, while being attractive to develop stable and efficient opto-electronic devices. 

\section*{Acknowledgment}
I.B.G acknowledges discussions with Cui Zhi-Hao on the implementation of the DSH functional. We acknowledge funding from the Agence Nationale pour la Recherche through the CPJ program and the SURFIN project (ANR-23-CE09-0001), the ALSATIAN project (ANR-23-CE50-0030), and the ANR under the France 2030 programme, MINOTAURE project (ANR-22-PETA-0015). M.Z. was funded by the European Union (project ULTRA-2DPK / HORIZON-MSCA-2022-PF-01 / Grant Agreement No. 101106654). Views and opinions expressed are however those of the authors only and do not necessarily reflect those of the European Union or the European Commission. Neither the European Union nor the  granting authority can be held responsible for them. We acknowledge computational resources from the EuroHPC Joint Undertaking and supercomputer LUMI [https://lumi-supercomputer.eu/], hosted by CSC (Finland) and the LUMI consortium through a EuroHPC Extreme Scale Access call, and access to the HPC resources of TGCC under the Allocation Grant No. 2022 - A0130907682 made by GENCI

\section*{Supporting Information Available}
Doubly screened dielectric dependent hybrid functional; 
Computational details;
Comparison of model dielectric functions in hybrid functionals;
Lattice constants and band gaps of 3D perovskites calculated using different functionals in Figure 2;
Band gaps of layered halide perovskites calculated using different functionals in Figure 3;
Band gaps of BA series in Figure 4;
Comparison of model dielectric functions for hybrid functionals in 3D perovskites;
Comparison of model dielectric functions for hybrid functionals in layered halide perovskites;
Convergence of dielectric constant and band gap with self-consistent DSH
iteration in 3D perovskites; Density of states of layered halide perovskites and BA$_2$MA$_{n-1}$Pb$_n$I$_{3n+1}$ ( n = 1, 2, 3).

\clearpage
\bibliography{main.bib}

\end{document}